\documentclass[pre,aps,showpacs,nofootinbib,twocolumn]{revtex4}
\usepackage{amssymb}

\usepackage{graphicx}

\begin{document}

\author{Miodrag L. Kuli\'{c}$^{1}$ and Igor M. Kuli\'{c}$^{2}$}
\title{\textbf{High-T}$_{c}$ \textbf{Superconductors with Antiferromagnetic Order: }%
\\
\textbf{Limitations on Spin-Fluctuation Pairing Mechanism}}

\begin{abstract}
The very intriguing antagonistic interplay of antiferromagnetism (AF) and
superconductivity (SC), recently discovered in high-temperature
superconductors, is studied in the framework of a microscopic theory. We
explain the surprisingly large increase of the magnetic Bragg peak intensity
$I_{Q}$ at $Q\sim (\pi ,\pi )$ in the magnetic field $H\ll H_{c2}$ at low
temperatures $0<T\ll T_{c},T_{AF}$ in $La_{2-x}Sr_{x}CuO_{4}$. Good
agreement with experimental results is found. The theory predicts large
anisotropy of the relative intensity $R_{Q}(H)=(I_{Q}(H)-I_{Q}(0))/I_{Q}(0)$%
, i.e. $R_{Q}(H\parallel c-axis)\gg R_{Q}(H\perp c-axis)$. The\ quantum
(T=0) phase diagram at H=0 is constructed. The theory also predicts: (i) the
magnetic field induced AF order in the SC state; (ii) small value for the
spin-fluctuation coupling constant $g<(0.025-0.05)$ $eV$. The latter gives
very small SC critical temperature $T_{c}(\ll 40$ $K)$, thus questioning the
spin-fluctuation mechanism of pairing in HTS oxides.
\end{abstract}

\pacs{75.10.Jm, 74.25.-q, 74.60.-w, 76.50.+g}

\address{$^{1}$Institut f\"{u}r Physik, Theoretische Physik II, \\
Universit\"{a}t Augsburg, \ 86135 Augsburg, Germany\\
$^{2}$Max-Planck-Institut f\"{u}r Polymerforschung, Theory Group\\
D-55021 Mainz, Germany} \maketitle

\newpage \textit{Introduction} - The antagonistic interplay of
superconductivity (SC) and antiferromagnetic (AF) order has been intensively
studied in compounds with regular distribution of magnetic ions, for
instance in $RERh_{4}B_{4},$ $REMo_{6}S_{8}(Se_{8})$, etc. - the systems
with localized rare earth (RE) magnetic moments, and $UPt_{3}$, URu$_{2}$Si$%
_{2}$ $UCu_{5}$, $UPd_{2}Al_{3}$ etc. - the systems with itinerant moments
\cite{BuBuKuPa}. The interest in the coexistence problem of the AF\ and SC
order has increased after the discovery of high-temperature superconductors
(HTS), in which the antiferromagnetic fluctuations and the tendency to the
AF order are pronounced. This fact implies inevitably several important
questions: (i) do the AF and SC order coexist in these materials and how
they influence each other? (ii) if the spin fluctuations are responsible for
pairing in HTS oxides, in which way are they affected by the AF order? (iii)
is the AF order compatible with other pairing mechanisms, such as the phonon
one \cite{KulRev}? There is a strong believe that by answering these
questions an important step will be done towards the understanding of the
pairing mechanism in HTS oxides, which is still under debate.

The first experimental evidence for the coexistence of the AF and SC order
in $La_{2-x}Sr_{x}CuO_{4}$ (for $x<0.15$) was reported in Ref.(\cite{Weiding}%
), which gave an impetus for theoretical studies of this problem in HTS\
oxides \cite{KuLiGoMe}. In the last few years careful neutron magnetic
scattering measurements on $La_{2}CuO_{4+y}$ \cite{Lee} , $%
La_{2-x}Sr_{x}CuO_{4}$ (for $x=0.12$, $T_{c}=12$ $K$, $T_{AF}\approx 25$ $K$%
) \cite{Katano}, $YBa_{2}Cu_{3}O_{6.5}$ (with $T_{c}=55$ $K$, $T_{AF}\approx
310$ $K$) \cite{Sidis}, $YBa_{2}(Cu_{1-y}Co_{y})_{3}O_{7+\delta }$ (with $%
y=0.013$ and $T_{c}=93$ $K$, $T_{AF}\approx 330$ $K$) \cite{Hodges}, have
given sufficient evidence for the coexistence of the AF and SC order. The
existence of the pronounced elastic magnetic Bragg peak at the wave vector $%
\mathbf{Q\sim Q}_{AF}=(\pi ,\pi )$ means that in these HTS\ oxides static AF
correlations are developed on the length scale $\xi _{m}\gtrsim 400$ \AA\
\cite{Lee}, \cite{Katano}. Since $\xi _{m}$ is much larger than the SC
coherence length in the a-b plane, $\xi _{ab}\sim (20-25)$ \AA ,\ this means
that these static AF correlations can be considered as a static AF order
which competes with superconductivity far outside the vortex core. The
experiments demonstrated that the effective magnetic moment $\mu $ is small,
i.e. $\mu \lesssim 0.1$ $\mu _{B}$ in $La_{2-x}Sr_{x}CuO_{4}$ and $\mu <0.05$
$\mu _{B}$ in $YBa_{2}(Cu_{1-y}Co_{y})_{3}O_{7+\delta }$.

The magnetic neutron scattering experiments performed on $%
La_{2-x}Sr_{x}CuO_{4}$ ($x=0.12$, $T_{c}=12$ $K$, $T_{AF}\approx 25$ K \cite
{Katano}) in magnetic field $H\sim 10$ $T$ and at $T=4$ $K$ show
surprisingly large increase of the elastic magnetic Bragg peak intensity $I_{%
\mathbf{Q}}$ by as much as $50$ $\%$ of that at $0$ $T$. This behaviour of $%
I_{\mathbf{Q}}(H)$ was confirmed recently on the same compound \cite{Lake} ($%
T_{c}=29$ $K$, $T_{AF}\approx 25$ $K$), where $I_{\mathbf{Q}}(H)$ $-I_{%
\mathbf{Q}}(0)\sim \alpha H\ln bH$ with an increase of $I_{\mathbf{Q}}(H)$
as much as $100\%$ at low $T$. Such a large increase of $I_{\mathbf{Q}}(H)$
(and its peculiar field dependence) exceeds by far the formally similar
effect in the magnetic superconductor $HoMo_{6}S_{8}$ \cite{Lynn} which
exhibits a sinusoidal magnetic order with $Q\ll Q_{AF}$ where $I_{\mathbf{Q}%
}(H)-I_{\mathbf{Q}}(0)\sim H^{2}$ for $H\lesssim 200$ $Oe$. The latter is
due to the suppression of the SC order in the presence of the exchange (or
spin-orbit) scattering, as it was explained in \cite{BuKu}.

The problem of the coexistence of the AF and SC order in high magnetic
fields was recently studied in the framework of the phenomenological
Ginzburg-Landau (GL) theory \cite{Demler} by assuming that the SC and AF
order parameters are small ($\mid \psi \mid \ll 1$, $m_{Q}\ll 1$). The
logarithmic behavior of $I_{\mathbf{Q}}(H)$ is found to be due to the
suppression of SC in the vortex state. However, the pronounced increase of $%
I_{\mathbf{Q}}(H)$ in the experiments on $La_{2-x}Sr_{x}CuO_{4}$ is realized
at very low temperatures ($T\ll T_{c}$) and in low fields $H\ll H_{c2}$,
where $H_{c2}(T\ll T_{c})\approx (50-80)$ $T$ . Thus the assumption that the
SC order parameter is small ($\mid \psi \mid \ll 1$ ) is not valid in these
experiments and a \textit{microscopic theory} of the problem is needed. The
latter is proposed and elaborated in this paper. The fact that $\mu (\sim
0.1-0.05)$ is small in the AFS systems of HTS oxides points to a \textit{%
weak itinerant antiferromagnetism}, which we also assume in the following -
the WIAF model. We shall calculate in this model the intensity $I_{Q}(%
\mathbf{H})$ in the coexistence phase for various field orientations. The
quantum ($T=0$ $K$) phase diagram is constructed and the condition for the
magnetic field induced AF order in the SC state is found.

\textit{WIAF model for AFS at }$T\ll T_{c}$ - The interplay of the AF and SC
order is studied in the framework of the weak coupling Hamiltonian
\begin{equation}
\hat{H}=\hat{H}_{0}(\vec{p}-\frac{e}{c}\mathbf{A})+\hat{H}%
_{Z}+\sum\limits_{i}gn_{i\uparrow }n_{i\downarrow }+\hat{H}_{BCS},
\label{Ham}
\end{equation}
where $\hat{H}_{Z}$ describes the Zeeman coupling of spins with $\mathbf{H}$
and $\mathbf{A}$ is the vector potential. The Hubbard term ($\sim g$) is
responsible for the weak itinerant AF order, while the $\hat{H}_{BCS}$
describes superconductivity with the SC order parameter $\Delta (\mathbf{k},%
\mathbf{R})=Y_{d}(\mathbf{k})\Delta _{0}(\mathbf{R})$, where $\mathbf{R}$ is
the mass centrum and $Y_{d}(\mathbf{k})$ is the d-wave function.

The study of the orbital effect of the magnetic field is mathematically
complicated since in fields $H_{c1}<H<H_{c2}$ the vortex state is realized
and $\Delta _{0}(\mathbf{R})$ varies in space. This variation occurs on a
much larger scale $\xi _{ab}$ than the atomic length $a$ (the length scale
of the AF order), i.e. $\xi _{ab},d_{v}\gg a$. The HTS materials are extreme
type II superconductors with the GL parameter $\kappa \gg 1$, and since the
AF order is developed on the large scale $\xi _{m}\gg \xi _{ab}$ it is
plausible to approximate $\Delta _{0}(\mathbf{R})$ by its average value over
the unit vortex cell with the intervortex distance $d_{v}$, i.e. we put $%
\Delta _{0}(\mathbf{R})\rightarrow \sqrt{<\Delta _{0}^{2}(\mathbf{R})>_{v}}$
similarly as it has been done in Ref. \cite{Demler}. In the case when $H\ll
H_{c2}$ one has \cite{Demler} $<\Delta _{0}^{2}(\mathbf{R})>_{v}\approx
\Delta _{0}^{2}\varphi ^{2}(h)$ with $\varphi (h)\approx \lbrack 1-(h/2)\ln
3/h]^{1/2}$ (for triangular vortex lattice) and $h=H/H_{c2}(\theta )$. The
upper critical field $H_{c2}(\theta )=H_{c2}^{c}/(\cos ^{2}\theta
+\varepsilon ^{2}\sin ^{2}\theta )^{1/2}$ depends on the angle $\theta $
between $\mathbf{H}$ and the crystall $c-axis$ where $\varepsilon
=H_{c2}^{c}/H_{c2}^{ab}$. So, the (orbital) effect of the vortex lattice on
superconductivity is accounted for by multiplying the SC order parameter by $%
\varphi (h)$. For further purposes we define the SC order parameter $\psi
(h)=\Delta (h)/\Delta _{0}\varphi (h)$, where $\Delta _{0}$ is the SC gap in
absence of the AF order and at $h=0$. We shall see that this physically
plausible approximation gives very good description of experiments on $%
La_{2-x}Sr_{x}CuO_{4}$.

\textit{Free-energy of AFS}\textbf{\ - }The magnetism in the WIAF model at $%
T\ll T_{AF}$ is treated in the Hartree-Fock approximation \cite{Moriya}. By
assuming that: (i) $h_{ex}(\equiv gm_{Q})\ll \Delta $, which is realized in $%
La_{2-x}Sr_{x}CuO_{4}$ - see below, (ii)\ $t=T/T_{c}\ll 1$ (iii) $\chi
_{sQ}/\chi _{nQ}(\Delta _{0})\lesssim 1$ - see below, the normalized Gibbs
free-energy $\tilde{G}(\equiv \chi _{n0}G)$ functional for the AFS in
magnetic field $\mathbf{h}$ has the form
\[
\tilde{G}=-\frac{a_{s}}{4}\varphi ^{2}(h)\psi ^{2}\ln \frac{e}{\psi ^{2}}+%
\frac{a}{2}m^{2}+\frac{b}{4}m^{4}-\mathbf{mh}\tilde{h}_{c2}
\]

\begin{equation}
-\frac{a_{_{Q}}}{2}m_{Q}^{2}+\frac{b_{_{Q}}}{4}m_{Q}^{4}+\frac{3}{4}%
bm^{2}m_{Q}^{2}+c\psi m^{2}+c_{Q}\psi m_{Q}^{2}.  \label{G}
\end{equation}
The first term describes the superconducting condensation energy (in absence
of the AF order) with $a_{s}=(N(0)\Delta _{0}{})^{2}$. The second and third
terms are due to the induced ferromagnetic (F) moment $m$ in the magnetic
field, where $a=1-2g\chi _{no}>0$ and $\chi _{no}=N(0)/2$ is the
free-electron normal state spin susceptibility at $\mathbf{q=0}$. The
parameter $b$ depends on the quasiparticle spectrum and for a weakly
anisotropic spectrum one has \cite{Moriya} $b=[(N^{\prime
}(0)/N^{2}(0))^{2}-N^{\prime \prime }(0)/3N^{3}(0)]/2$, where $N(0)$, $%
N^{\prime }(0)$, $N^{\prime \prime }(0)$ are the density of states and their
derivatives at the Fermi surface, respectively. The fourth term is the
energy of the magnetic moment in the field $\mathbf{h=H}/H_{c2}$ where $%
\tilde{h}_{c2}=N(0)(\mu _{B}H_{c2})$. The weak itinerant AF order (in
absence of the SC order) is described by the first two terms in the second
line in $Eq.(\ref{G})$, where $m_{Q}$ is the AF order parameter, $%
a_{_{Q}}=(2\chi _{nQ}g-1)\chi _{no}/\chi _{nQ}$ and $\chi _{nQ}$ is the
free-electron normal state spin susceptibility at $\mathbf{Q}\sim \mathbf{Q}%
_{AF}=(\pi ,\pi )$, while $b_{Q}=3b/8$ as in \cite{BuKu}. The third term in
the second line in $Eq.(\ref{G})$ is due to the interaction of the F and AF
components of the magnetic moment, while the terms proportional to $c\approx
\varphi (h)(\chi _{n0}/\chi _{s0}(\Delta _{0})/2$ and $c_{Q}\approx r\varphi
(h)(\chi _{nQ}/\chi _{sQ}(\Delta _{0})-1)/2$ are due to the competition of
the F and SC order, and of the AF and SC order, respectively. Here, $r=\chi
_{n0}/\chi _{nQ}$ where $\chi _{s0}(\Delta _{0})$ and $\chi _{sQ}(\Delta
_{0})$ are the free-electron spin susceptibility in the SC state with $\psi
=1$ at the wave vectors $\mathbf{0}$ and $\mathbf{Q}$, respectively. In the
case of a clean ($l\gg \xi _{ab}$) singlet d-wave superconductor, which we
assume here, $\chi _{s0}$ is strongly suppressed, i.e. $\chi _{s0}/\chi
_{n0}\approx 1.4(T/\Delta )\ll 1$ at $T\ll T_{c}$. The susceptibility $\chi
_{sQ}$ at $Q\approx (\pi ,\pi )$ is much less suppressed than at $q=0$, i.e.
$\chi _{sQ}(\Delta _{0})\approx \chi _{nQ}(1-\Delta _{0}/E_{Q})$ for $\Delta
_{0}/E_{Q}\ll 1$, where $E_{Q}$ depends on the electronic spectrum. (For a
parabolic spectrum $E_{Q}\approx \hbar v_{F}Q\gg $ $\Delta .)$. We stress,
that in the proposed microscopic theory the term in the Gibbs free-energy
which describes the interaction of the AF and SC order has the form $\tilde{G%
}_{int}^{micro}\sim \mid \psi \mid m_{Q}^{2}$ at $T\ll T_{c}$, contrary to
the GL expression $\tilde{G}_{int}^{GL}\sim \mid \psi \mid ^{2}m_{Q}^{2}$
which holds at $T\lesssim T_{c}$ and which is assumed in Ref. \cite{Demler}.

\begin{figure}[tbp]
\includegraphics*[width=8cm]{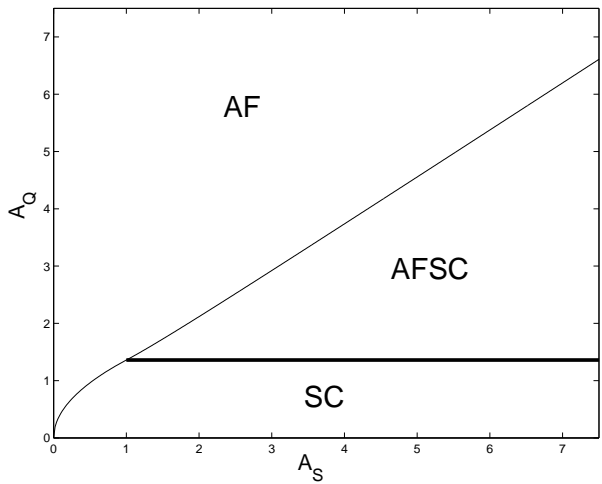}
\caption{Quantum phase diagram ($A_{s},A_{Q}$) of the AFS system for $%
c_{Q}=0.01$ and $b_{Q}=1/6$. The thin and thick lines are the first and
second order transition lines, respectively. $AF$ - ($m_{Q}\neq 0$, $\protect%
\psi =0$), $SC$ - ($m_{Q}=0$, $\protect\psi \neq 0$) and $AFSC$ - ($%
m_{Q}\neq 0$, $\protect\psi \neq 0$).}
\end{figure}

\textit{Phase diagram} \textit{at }$h=0$ - By minimizing $\tilde{G}(\psi
,m_{Q},m;h=0,T=0)$ the quantum phase diagram - shown in $Fig.1$, is
calculated by taking physically plausible values for $c_{Q}=0.01$, $%
b_{Q}=1/6 $. The latter values can also explain $I_{Q}(\mathbf{H})$ for $%
H\neq 0$ - see below. The phase diagram is studied in terms of the
parameters $a_{s}(\equiv 2.4\times 10^{-3}$ $A_{s})$ and $a_{Q}(\equiv
(4/e)\times 10^{-2}$ $A_{Q})$ which contains the following transition lines:
(\textit{i}) a \textit{first order} transition line AF$\longleftrightarrow $%
SC given by $A_{Q}=(e/2)(A_{s})^{1/2}$ for $A_{s}<1$; (\textit{ii}) a
\textit{first order} transition line AFSC$\longleftrightarrow $AF described
by $A_{Q}=(\sqrt{e}/2)A_{s}\exp (1/2A_{s})$ for $A_{s}>1$ and (\textit{iii})
a AFSC$\longleftrightarrow $SC \textit{second order} line $A_{Q}=e/2$ on
which the AF order starts growing from zero.

\textit{Magnetic Bragg peak intensity }$I_{\mathbf{Q}}(h)$ - Since $I_{%
\mathbf{Q}}(h)\sim m_{Q}^{2}$ we first calculate $m_{Q}^{2}$ in the case
when $\mathbf{H}\parallel c-axis$ ($\mathbf{m}_{Q}\perp c-axis$) by taking $%
N(0)\approx 1/100$ $meV$ \cite{Loram}, $\Delta _{0}\approx 10$ $meV$ (in $%
La_{2-x}Sr_{x}CuO_{4}$) which gives $a_{s}\approx 10^{-2}$. Furthermore, in
clean $d$-wave SC one has $c\gtrsim 10$ at low $T\lesssim 1$ $K$, while $%
b\approx 0.5$ follows from the WIAF theory \cite{Moriya}. We take $r\approx
0.25$ in accordance with the model calculations of the quasiparticle
spectrum in HTS oxides \cite{Bulut}. Since in low-T$_{c}$ AFS systems one
has $\beta \ll 1$ we take $\beta \approx 0.1$, which is also in accordance
with the model calculations \cite{Bulut} for $\chi _{nQ}^{0}-\chi _{sQ}^{0}$.

\begin{figure}[tbp]
\includegraphics*[width=8cm]{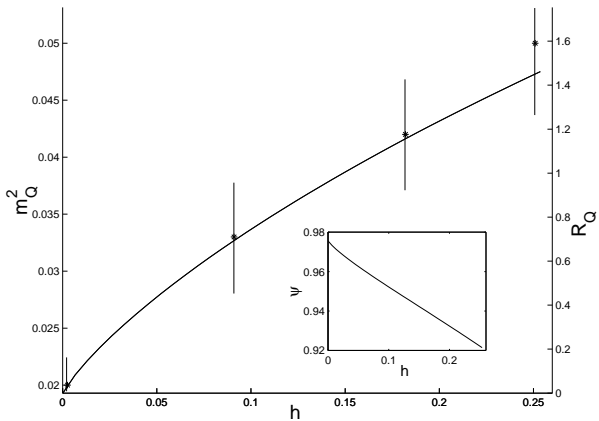}
\caption{Case $H\parallel c-axis$. Magnetic field\ ($h=H/H_{c2}^{c}$)
dependence of $m_{Q}^{2}$, $R_{Q}$ and $\protect\psi $ (inset) for $%
a_{s}=0.01$, $a_{Q}=0.028$ $b=0.5$, $t=0.1$, $r=0.25$, $\protect\beta =0.1$
and $\tilde{h}_{c2}=0.1$. Points and errorbars are experimental results
\protect\cite{Lake} by assuming that $H_{c2}^{c}\approx 50$ $T$.}
\end{figure}

The calculated functions $m_{Q}^{2}(h)$, $R_{Q}(h)$ and $\psi (h)$, for $%
a_{s}=0.01$, $b=0.5$, $t=0.1$, $r=0.25$, $\beta =0.1$, $a_{Q}=0.028$, are
shown in $Fig.2$. The relative intensity $R_{Q}(h)$ behaves like $%
R_{Q}(h)\approx A(1-\varphi (h))$, while the quadratic term $\sim
Bh^{2}<0.01 $ (which dominates in $HoMo_{6}S_{8}$) is negligible. By
assuming $H_{c2}^{c}\approx 50$ $T$ (at $T\ll T_{c}$), and having in mind
the large experimental errorbar, one obtains very good \textit{quantitative
agreement }of the proposed theory and experimental results \cite{Lake} on $%
La_{2-x}Sr_{x}CuO_{4}$, as it is seen in $Fig.2$. Note, that in this
compound (and for $h=0$) the bare AF order parameter $m_{Q0}$ is drastically
suppressed by superconductivity, i.e. $m_{Q}^{2}=m_{Q0}^{2}-2c_{Q}\psi \ll
m_{Q0}^{2}$. We stress also that the AF order is not only realized inside
the vortex core ($0<R_{cor}<\xi _{ab}$) but also in the bulk region given by
$\xi _{ab}<R_{bulk}<d_{v}$ ($d_{v}\gg \xi _{ab}$ at $H\ll H_{c2}$), which is
in accordance with the experiments on $La_{2-x}Sr_{x}CuO_{4}$ \cite{Katano}.
The calculations also show that in the narrow region $h=0.3-0.4$ there is a
steep decrease of $\psi $ while $m_{Q}$ grows sharply (similarly as in $%
Fig.4 $ for $\mathbf{H\perp c}$-axis).

Second, in the field $\mathbf{H}\perp c$-axis and for $H_{c2}^{ab}(T=1$ $%
K)\sim 10H_{c2}^{c}(T=1$ $K)$ one gets a small value for $R_{Q}(h)\sim 0.01$
at $H\sim 10$ $T$, while for $h=0.07-0.08$, $m_{Q}^{2}$ increases sharply
and $\psi $ drops to zero - $Fig.3$. Due to this large field anisotropy one
expects that in polycrystalline samples of $La_{2-x}Sr_{x}CuO_{4}$
superconductivity vanishes in fields $20$ $T<H<40$ $T$ in the percolation
process \cite{KicIr}.

\begin{figure}[tbp]
\includegraphics*[width=8cm]{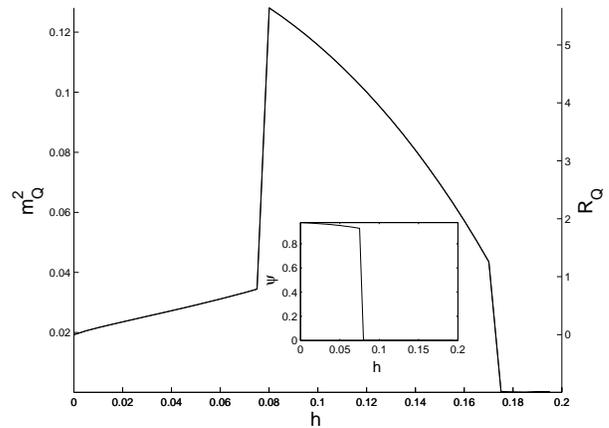}
\caption{Case $H\perp c-axis$. Magnetic field ($h=H/H_{c2}^{ab}$) dependence
of $m_{Q}^{2}$, $R_{Q}$ and $\protect\psi $ (inset) for parameters in $Fig.2$
and for $h_{c2}^{ab}=10h_{c2}^{c}$.}
\end{figure}

A very interesting situation is realized if $c_{Q}(\ll 1)$ is increased by $%
20\%$ in which case the AF order is \textit{induced} ($m_{Q}\neq 0$) in
magnetic fields $h>0.07$ - see $Fig.4$. This sensitivity of the AF order on $%
c_{Q}$ may explain its absence in some HTS oxides.

\begin{figure}[tbp]
\includegraphics*[width=8cm]{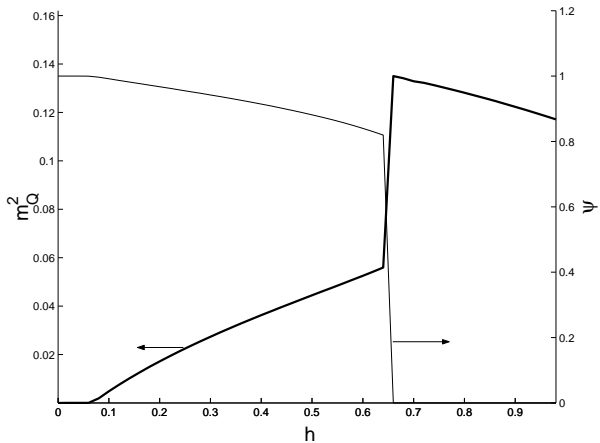}
\caption{Induced $AF$ order for $H\parallel c-axis$. Magnetic field\ ($%
h=H/H_{c2}^{c}$) dependence of $m_{Q}^{2}$ and $\protect\psi $ for $%
a_{s}=0.01 $, $a_{Q}=0.028$ $b=0.5 $, $t=0.1$, $r=0.25$, $\protect\beta %
=0.12 $ and $\tilde{h}_{c2}^{c}=0.1$.}
\end{figure}

\textit{Discussion - }The microscopic theory proposed here predicts that the
interplay of the SC and AF order is controlled by the interaction parameter $%
c_{Q}$. The experiments on $YBa_{2}Cu_{3}O_{7-x}$ imply that $c_{Q}(\ll
0.01) $ is very small, and therefore the SC order coexists with the AF
itinerant magnetism (with $m_{Q}\approx m_{Q0}\ll 1$) more easily. One
expects in this case, that the AF order is weakly affected by the magnetic
field, contrary to the case of $La_{2-x}Sr_{x}CuO_{4}$ discussed above.

Since the WIAF model for the AF order in HTS oxides explains the intensity $%
I_{Q}(\mathbf{h})$ very well, it is a challenge to estimate the
spin-fluctuation coupling constant $g$ entering $a_{Q}$. For $a_{Q}\sim
0.028 $ (if $b\sim 0.5$), which fits $I_{Q}(\mathbf{h})$ very well, or even
for $a_{Q}\sim 1$ (but $b>1$), one obtains $2\chi _{nQ}^{0}g\lesssim (1-2)$,
which gives $g<(0.25-0.5)/N(0)$. For $N(0)\sim (100$ $meV)^{-1}$ one obtains
$g<(25-50)$ $meV$, i. e. $g\ll g^{sf}\approx 0.65$ $eV$, where $g^{sf}$
being the (very large) coupling constant assumed in theories based on the
spin-fluctuation pairing mechanism in HTS oxides in order to explain $%
T_{c}\sim 100$ $K$ - see \cite{PinesRev}. The obtained small value
of $g(\ll g^{sf})$ in the AFS materials of the HTS oxides tells us
that $T_{c}$, which is due to the spin-fluctuation (SF)
interaction, is very small, in agreement with the value
$g^{sf}\approx 15$ $meV$ obtained recently by analyzing the
resonance peak effects in the SC state of HTS oxides
\cite{Kivelson}. Serious arguments against the SF pairing
mechanism were given in Ref. \cite {KulRev}, which are based on
the experimental fact that by increasing doping from
$YBa_{2}Cu_{3}O_{6.92}$ to $YBa_{2}Cu_{3}O_{6.97}$, there is a
large redistribution (over $\omega $) of the SF spectral function
$Im \chi (Q,\omega ),$ while $T_{c}$ is changed only slightly,
i.e. $T_{c}=91$ $K$ in $YBa_{2}Cu_{3}O_{6.92}$ and $T_{c}=92.5$
$K$ in $YBa_{2}Cu_{3}O_{6.97}$. This result can be explained by
invoking very small coupling between conduction electrons and
spin-fluctuations, i.e. $g^{sf}\ll 0.65$ $eV$.
Note, that the small $g$ value gives that $h_{ex}\sim (2-5)$ $meV$ in $%
La_{2-x}Sr_{x}CuO_{4}$, i.e. the condition $h_{ex}\ll \Delta _{0}$ used in
the calculation is fulfilled.

In conclusion, the proposed microscopic theory for the interplay of the AF
and SC order in HTS oxides is able to explain quantitatively the magnetic
field dependence of the magnetic Bragg scattering intensity $I_{Q}(\mathbf{H}%
)$ in $La_{2-x}Sr_{x}CuO_{4}$ for $x<0.15$. A large anisotropy of the
relative intensity $R_{Q}(\mathbf{H)}$ is predicted, i.e. $R_{Q}(\mathbf{%
H\parallel c}$-axis$)\gg R_{Q}(\mathbf{H\perp c}$-axis$)$, which implies
that in polycrystalline samples of $La_{2-x}Sr_{x}CuO_{4}$ superconductivity
disappears in the percolation process in fields, $20$ $T<H<40$ $T$ (for $%
H_{c2}^{c}\sim 50$ $T$). In the quantum phase diagram (for $h=0$) one second
order and two first order transition lines meet at the same point. The
theory also predicts: (i) that the AF order parameter is strongly
renormalized by the SC order in $La_{2-x}Sr_{x}CuO_{4}$; (ii) that in the SC
state of some HTS oxides the AF order can be induced by a finite magnetic
field; (iii) that the small value for the spin-fluctuation (SF) coupling
constant, $g<(0.025-0.05)$ $eV$, makes the SF mechanism ineffective in
producing high $T_{c}$ in HTS oxides.

M. L. K. acknowledges the support of the Deutsche Forschungsgemeinschaft (%
\textit{DFG}) through the \textit{SFB 484} project at University of Augsburg
and the support of Ulrich Eckern and Franz Wegner.

\end{document}